\documentclass{jfm}

\usepackage{graphicx}
\usepackage{newtxtext}
\usepackage{newtxmath}
\usepackage{natbib}
\usepackage{hyperref}
\hypersetup{
    colorlinks = true,
    urlcolor   = blue,
    citecolor  = black,
}

\newcommand{\RomanNumeralCaps}[1]
\usepackage{amsmath}
\usepackage{braket}
\usepackage{psfrag,epsfig}
\usepackage{subfigure}
\usepackage[percent]{overpic}
\usepackage{float}
\usepackage{color}
\usepackage{tikz}
\usetikzlibrary{arrows.meta,bending,positioning,intersections}
\usepackage{pgfplots}
\usepackage[all]{xy}
\graphicspath{{figures/}}

\title{Hyperbolic free-surface jets}

\author{Andrew Wilkinson\aff{1},
Michael A. Morgan\aff{2},
\and Michael Wilkinson\aff{1}
\corresp{\email{m.wilkinson@open.ac.uk}},
}

\affiliation{\aff{1}School of Mathematics and Statistics, The Open University, 
Walton Hall, Milton Keynes MK7 6AA, UK
\aff{2}Department of Physics, Seattle University, Seattle, WA98122, USA}

\begin{document}
\maketitle

\begin{abstract}
If a body of inviscid fluid is disturbed, it will typically eject a \emph{jet} 
of fluid. If the effects of gravity and surface tension are negligible, these 
jets travel in straight lines, with the tips approaching a constant velocity.
It has been observed that these jets can have a broad base, tapering 
progressively toward the tip, but the mathematical form of their profile has not been 
successfully analysed in earlier works. In this paper, we describe the simplest 
case, in two dimensions: an infinitely deep body of inviscid fluid, with no surface tension 
or gravitational forces acting, responds to an impulsive disturbance. 
We find that, contrary to some earlier suggestions, the jet has a 
hyperbolic profile (away from its tip and its base).
\end{abstract}

\begin{keywords}
%Authors should not enter keywords on the manuscript, as these must be chosen by the author during the online submission process and will then be added during the typesetting process (see \href{https://www.cambridge.org/core/journals/journal-of-fluid-mechanics/information/list-of-keywords}{Keyword PDF} for the full list).  Other classifications will be added at the same time.
\end{keywords}

%{\bf MSC Codes }  {\it(Optional)} Please enter your MSC Codes here
\section{Introduction}
\label{sec: 1}

Free surface problems in fluid dynamics are usually challenging. The 
case of incompressible, inviscid flow, in situations where surface tension may also be 
neglected, is of particular importance for analysing ocean waves. 
If a body of inviscid fluid is disturbed, at least some fluid elements will
be set on trajectories which will take them outside the initial boundary. 
Once a region of fluid starts to escape from the bulk, it may form a slender \emph{jet}.
As the jet extends, pressure gradients reduce, and the fluid elements which enter the 
jet move ballistically.

In many problems, such as ocean waves and splashes of objects dropped into the sea, 
gravity will play an important role. However, it is of interest to understand the 
solution of this problem in contexts where there is no gravity, for two reasons.
Firstly, a clear understanding of the case without gravity is a foundation for understanding 
cases where gravity is important. Secondly, in astrophysical processes, bodies 
of fluid could be subjected to strong tidal forces during a transient encounter, 
before freezing into an exotically extended shape. A possible example is the 
\emph{Oumuamua} object \citep{Mee+17}, an asteroid size object of interstellar origin, 
which appeared to have a highly elongated structure.

In this paper we consider the simplest possible non-trivial case. There is an infinitely 
deep body of incompressible, inviscid fluid, which is initially stationary, 
with a flat interface, and the complementary half-space 
is assumed occupied by a gas of negligible density. The interface has 
zero surface tension. The system is subject to an impulsive disturbance, 
such that at $t=0$, the velocity is instantaneously 
$\mbox{\boldmath$u$}=\mbox{\boldmath$\nabla$}\phi$. 
Only the two-dimensional version of this problem will be addressed here.
This problem is related to, but distinct from, splashback from dropped 
objects impacting at high Reynolds number. This latter problem has multiple complications, 
involving gravity, vorticity generated by shear at the surface of the projectile, 
cavitation, and turbulent dissipation (see, e.g. \cite{O'B+95}). 

The natural questions are: From which points on the surface (if any) do the jets emerge? 
What is the velocity of the tip of the jet? How does the shape of the jet evolve?
How much material is ejected in the jet? Or, as a possible alternative, 
does all of the fluid eventually find its way into the jet?

We have found surprisingly little literature on this foundational question. 
An exactly solvable case, that of a circular body subjected to a simple extensional 
flow, was discovered by \cite{Dir60}, who showed that the circle deforms 
into an ellipse with increasing aspect ratio. Further developments of this approach 
are discussed in \cite{Lon72}. The more general case which we consider was 
previously discussed by \cite{Lon01a}, who 
made the intriguing suggestion that the jet tip may have a cusped profile, described by a \lq canonical form', 
which is somewhat reminiscent of the normal form for a cusp in \lq catastrophe theory'
(reviewed in \cite{Pos+96}). We find that the form of the 
jet profile is different from that proposed by Longuet-Higgins, the width profile being 
a hyperbola, rather than a 3/2 power-law cusp.  Some closely related 
ideas are presented in further works by Longuet-Higgins and collaborators, see
\cite{Lon01b,Lon+01a,Lon+01b}. 

We remark that there are a disparate set of phenomena in fluid flow which 
are described as \lq jets'. A well-known 
review \cite{Egg+08}  defines a \lq jet' as a \lq collimated stream of matter having 
a more-or-less columnar shape'. We shall use the term \emph{jet} despite the flow being 
tapering rather than columnar.
We are also aware of two other approaches to the problem of the motion of a two-dimensional  
perfect fluid with a free surface. The first of these, as exemplified by \cite{KSZ+94}, 
\cite{ZKD+96a} and \cite{DKSZ+96b}, uses a Hamiltonian formalism with the 
surface displacement and surface potential as conjugate fields, combined with 
a conformal mapping to the half-plane. However, this approach is perturbative 
in small displacements and angles. 
Examples of the second approach are \cite{KZ+14} and \cite{KZ+24}. These papers 
give exact solutions contrived to obey an auxiliary equation which keeps the 
free surface moving with a constant velocity. Neither of these approaches 
were useful for our problem, because we consider a non-perturbative flow of a rather general form.

For potential flow, $\mbox{\boldmath$u$}=\mbox{\boldmath$\nabla$}\phi$. 
If the fluid is incompressible, we have $\nabla^2\phi=0$. If the flow is inviscid, 
the potential and pressure $p$ are related by the time-dependent Bernouilli equation
\begin{equation}
\label{eq: 1.1}
\frac{\partial \phi}{\partial t}+\frac{1}{2}(\mbox{\boldmath$\nabla$}\phi)^2+\frac{1}{\rho}p=\Phi(t)
\ .
\end{equation}   
In the following, we take $\rho=1$. 
We assume that the initial boundary is $y=0$, and that the pressure is $p=0$ at the surface.
The fluid is disturbed by an impulse at $t=0$, which will be assumed to be periodic in $x$. 
To avoid unnecessary complications, we discuss the following simple form for the 
initial velocity potential:
\begin{equation}
\label{eq: 1.2}
\phi(x,y,0)=\cos(x)\exp(y)+\epsilon \cos (2x)\exp(2y)
\end{equation}
and we assume $|\epsilon| <1/4$. 
The initial velocity field is 
\begin{equation}
\label{eq: 1.3}
\mbox{\boldmath$u$}=\left(-\sin(x)\exp(y)-2\epsilon\sin(2x)\exp(2y),\cos(x)\exp(y)
+2\epsilon\cos(2x)\exp(2y)\right)
 \ .
\end{equation}
Note that there is a converging fixed point of the $x$-coordinate of fluid 
elements on the surface at $x=0$, although the surface itself moves in the $y$-direction,  
initially rising with speed $u_y=1+2\epsilon$ at that point. Similarly there is a diverging 
fixed point of the surface flow at $x=\pm \pi$ where the initial vertical speed 
$u_y=-1+2\epsilon$ is downward.  

We expect that the upward, converging fixed point of the surface flow becomes the tip of a 
jet moving vertically upwards. As material in the vicinity of this point escapes from the surface, 
the velocity gradients decrease, and (\ref{eq: 1.1}) indicates that pressure gradients approach zero, 
so that the escaping fluid elements move ballistically (i.e. with constant velocity) at large time. 
In particular, the speed of the tip is expected to approach a constant value 
$v^\ast$ as $t\to \infty$. The diverging fixed point of the surface flow is expected to become the 
lowest point of a trough in the surface. It is a non-trivial task to determine the form of the jet, which in this paper 
will be described by estimating the surface profile $y_{\rm s}(x,t)$.

\begin{figure}
\centering
\includegraphics[width=10cm]{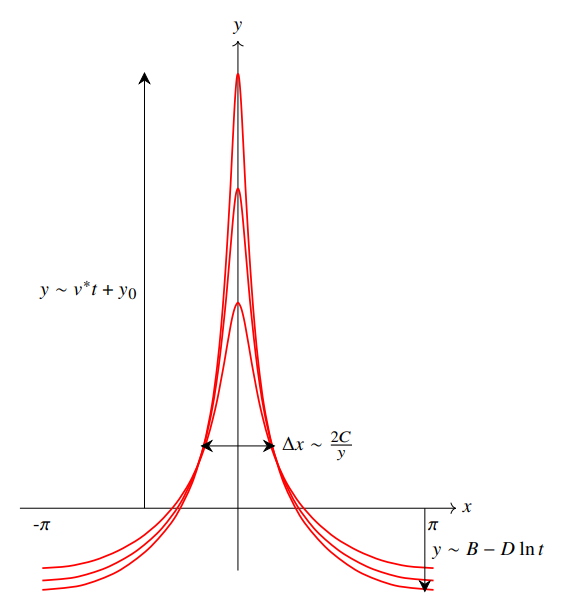}
\caption{
Schematic illustration of the form of the jet. The tip of the jet extends with a constant speed. 
The profile of the base of the jet approaches a curve which is independent of time, apart
from a downward displacement proportional to $\ln t$. The body of the jet has a hyperbolic form:
$y\Delta x=2C$, where $C$ is a constant.
}
\label{fig: 1}
\end{figure}

We were not able to find an exact solution to determining the surface profile. 
Figure \ref{fig: 1} is a schematic illustration of the form of the jet at different times. 
After a brief initial transient, the position of the tip is
approximated by 
\begin{equation}
\label{eq: 1.7}
y_{\rm s}(0,t)\sim v^\ast t+y_0
\end{equation}
where both the tip speed $v^\ast$ and the offset $y_0$ must be determined numerically 
(by fitting the empirically determined height of the jet at large $t$).
The principal results of our analysis are a theory for the width of the jet, and 
for the depth of the trough. We argue that, as a function of height $y$ above the 
original surface, the \emph{full} width of the jet is 
\begin{equation}
\label{eq: 1.4}
\Delta x\sim \frac{2C}{y}
\end{equation}
for some constant $C$.
This relation applies at sufficiently large times, with $y$ satisfying $y\ll v^\ast t$ 
and $y\gg 1$.
At the base of the jet, we argue that the lowest point of the trough is, asymptotically 
as $t\to \infty$,
\begin{equation}
\label{eq: 1.5}
y_{\rm s}(\pm \pi)\sim B-D\ln(t)
\end{equation}
for some constants $B$, and $D >0$. We also argue that, as $t\to \infty$, the base of 
the jet retains its form, so that
\begin{equation}
\label{eq: 1.6}
y_{\rm s}(x,t)\sim F(x)+B-D\ln(t)
\end{equation}
for some function $F(x)$, satisfying $F(\pm \pi)=0$ and $F(x)\to \infty$ as $x\to 0$. 
This approximation is accurate in the limit as $t\to \infty$, except in a 
small (and shrinking) interval surrounding the tip of the jet at $x=0$.

Section \ref{sec: 2} will discuss some general principles which form the basis 
for our analysis. Section \ref{sec: 3} will present the argument which supports 
equation (\ref{eq: 1.4}). 
In section \ref{sec: 4} we derive an expression for the surface profile $y_s(x,t)$ valid in the 
vicinity of the jet tip, and use conservation of energy to relate the tip velocity $v^\ast$ to the coefficient $C$.
Section \ref{sec: 5} discusses the form of the base of 
the jet, justifying equations (\ref{eq: 1.5}) and (\ref{eq: 1.6}). Our investigations were 
informed by numerical simulations of the flow, based upon the method of \cite{Lon+76}. 
We had to make significant modifications to their method, which are described in Appendix A. 
Section \ref{sec: 6} uses an interpolation to combine our two approximate solutions 
(which describe, respectively, the jet and the base region), showing excellent agreement
with numerical simulations. This final section also presents a brief conclusion.

\section{Basic principles}
\label{sec: 2}

\subsection{Equation of motion for boundary}
\label{sec: 2.1}

Analysing the motion of free surfaces in inertia-dominated flows is a challenging problem. 
A very significant advance was introduced in \cite{Lon+76}, who gave equations of motion
for fluid elements on the boundary, and for values of the velocity potential on the boundary. 
These equations are to be supplemented with a procedure for finding a harmonic function 
which represents the velocity potential in the interior of the region. In \cite{Lon+76}, a 
boundary integral method is used to determine this harmonic function. In this paper, we use 
analytical approximations instead (primarily, using the \lq slender body' concept).

\cite{Lon+76} considered a two-dimensional flow, in a region with Cartesian coordinates $(x,y)$. 
They show that the equations of motion for the boundary are
\begin{equation}
\label{eq: 2.1}
\frac{{\rm D}x}{{\rm D}t}=\frac{\partial \phi}{\partial x}
\ ,\ \ \  
\frac{{\rm D}y}{{\rm D}t}=\frac{\partial \phi}{\partial y}
\ ,\ \ \  
\frac{{\rm D}\phi}{{\rm D}t}=\frac{1}{2}(\mbox{\boldmath$\nabla$}\phi)^2
\ .
\end{equation}
The potential $\phi(x,y,t)$ is a harmonic function, which is determined by the evolution 
of $\phi$ on the boundary. (It is necessary to consider the potential in the interior region 
in order to determine the component of $\mbox{\boldmath$\nabla$}\phi$ normal 
to the boundary, which is required in (\ref{eq: 2.1})).
Their paper also introduced a conformal transformation from the $z=x+{\rm i}y$ variable, 
to $w=u+{\rm i}v$:
\begin{equation}
\label{eq: 2.2}
w=u+{\rm i}v=r\exp({\rm i}\theta)=\exp[-{\rm i}(x+{\rm i}y)]=\exp(-{\rm i}z)
\ .
\end{equation}
The potential $\phi$ remains a harmonic function when expressed in terms 
of the $(u,v)$ variables, but the equations of motion for the boundary are transformed. 
We shall require them in both Cartesian and polar coordinates of the $w$-plane. 
The polar coordinate form was given in \cite{Lon+76}:
\begin{equation}
\label{eq: 2.3}
\frac{{\rm D}r}{{\rm D}t}=r^2\frac{\partial \phi}{\partial r}
\ ,\ \ \ 
\frac{{\rm D}\theta}{{\rm D}t}=\frac{\partial \phi}{\partial \theta}
\end{equation}
and for evolution of potential on the boundary
\begin{equation}
\label{eq: 2.4}
\frac{{\rm D}\phi}{{\rm D}t}=\frac{1}{2}\left[\left(r\frac{\partial \phi}{\partial r}\right)^2
+\left(\frac{\partial \phi}{\partial \theta}\right)^2\right]
\ .
\end{equation}
In terms of the $(u,v)$ variables, these equations of motion are
\begin{eqnarray}
\label{eq: 2.5}
\frac{{\rm D}u}{{\rm D}t}&=&(u^2+v^2)\frac{\partial \phi}{\partial u}
\nonumber \\
\frac{{\rm D}v}{{\rm D}t}&=&(u^2+v^2)\frac{\partial \phi}{\partial v}
\end{eqnarray}
\begin{equation}
\label{eq: 2.6}
\frac{{\rm D}\phi}{{\rm D}t}=\frac{(u^2+v^2)}{2}
\left[\left(\frac{\partial \phi}{\partial u}\right)^2+\left(\frac{\partial \phi}{\partial v}\right)^2\right]
\ .
\end{equation}
In the $w$-plane, the initial condition is a unit circle centred on the origin, and the 
potential at $t=0$ is $\phi(u,v)={\rm Re}(w+\epsilon w^2)$.

\subsection{Global conservation laws}
\label{sec: 2.2}

The initial kinetic energy 
\begin{equation}
\label{eq: 2.7}
E=\frac{1}{2}\int_{-\pi}^\pi {\rm d}x\int_{-\infty}^0{\rm d}y\ 
[\mbox{\boldmath$\nabla$}\phi]^2=\frac{\pi}{2} [1+2\epsilon^2]
\end{equation}
is conserved, as is the volume, implying 
\begin{equation}
\label{eq: 2.8}
{\cal V}=\int_{-\pi}^\pi {\rm d}x\ y_{\rm s}(x,t)=0
\end{equation}
where $y_{\rm s}(x,t)$ is the surface height, initially $y_{\rm s}(x,0)=0$.  
We do not use conservation of momentum and angular momentum 
explicitly. 

\subsection{A local conservation principle}
\label{sec: 2.3}

Consider the motion of material points on the surface.
Because the surface has constant pressure, there is no acceleration
of fluid elements along lines tangent to the surface.
Pressure gradients in the direction normal to the boundary can accelerate
fluid elements normal to the boundary.
Changes of the relative speed of two elements on the surface
can, therefore, only occur if the surface is curved, so that the normals point
in different directions.

Consider the relevance of these observations to the motion of 
points on the surface which start in the vicinity of the unstable fixed point 
of the surface flow, at $x=\pm \pi$. A point which is displaced from this fixed point 
by a distance $\delta x$ has horizontal speed (to leading order in $\delta x$)
\begin{equation}
\label{eq: 2.9}
u_x=(1-4\epsilon)\delta x
\ .
\end{equation}
The point continues to move with the same horizontal speed until the slope of the boundary 
is significantly different from zero.

\subsection{Velocity gradient}
\label{sec: 2.4}

While not essential to our arguments, it is informative to consider the velocity gradient matrix
\begin{equation}
\label{eq: 2.10}
{\bf A}=\left(\begin{array}{cc}
\partial^2_{xx}\phi & \partial^2_{xy}\phi \cr
\partial^2_{xy}\phi & \partial^2_{yy}\phi
\end{array}\right)
\ .
\end{equation}
In two-dimensional potential flow the velocity gradient matrix is a traceless 
symmetric matrix. It follows that the eigenvectors are orthogonal, and the eigenvalues
are both real, and that their sum is zero. 

The equation of motion for ${\bf A}$ is (in the case of an inviscid fluid)
\begin{equation}
\label{eq: 2.11}
\frac{{\rm D}{\bf A}}{{\rm D}t}+{\bf A}^2=-\frac{1}{\rho}{\bf H}
\end{equation}
where ${\bf H}$ is the Hessian matrix of the pressure (with elements
$H_{ij}=\partial^2_{ij}p$) (see \cite{Can92}). 

At long times, we expect pressures, and hence the pressure Hessian, to approach zero.
If ${\bf H}$ is negligible, equation (\ref{eq: 2.11}) indicates that $A_{ij}\sim 1/t$ as 
$t\to \infty$. This, in turn, would indicate that we might write
\begin{equation}
\label{eq: 2.12}
\phi(x,y,t)\sim \frac{1}{t}\psi(x,y)
\end{equation}
in the limit as $t\to \infty$, where $\psi(x,y)$ is a harmonic function. Approximations 
of this form will figure in our analysis.

\subsection{Physical principles of the solution}
\label{sec: 2.5}

Immediately after the impulse is applied at $t=0$, the pressure in the fluid can be determined. 
(For completeness, the calculation is discussed in Appendix B).
We find that the pressure at depth increases instantaneously, implying an upward
acceleration of fluid elements close to the surface. This must be a transient effect, 
(otherwise it would accelerate them to the extent that energy would not be conserved).
The pressure field immediately following the initial impulse can, therefore, be 
thought of as a short-lived pulse which imparts a secondary impulse to the fluid elements.
We were not able to make a precise statement about this secondary impulse, 
so that the initial response of the system must be determined by a numerical 
calculation.

\begin{figure}
\centering
\includegraphics[width=14cm]{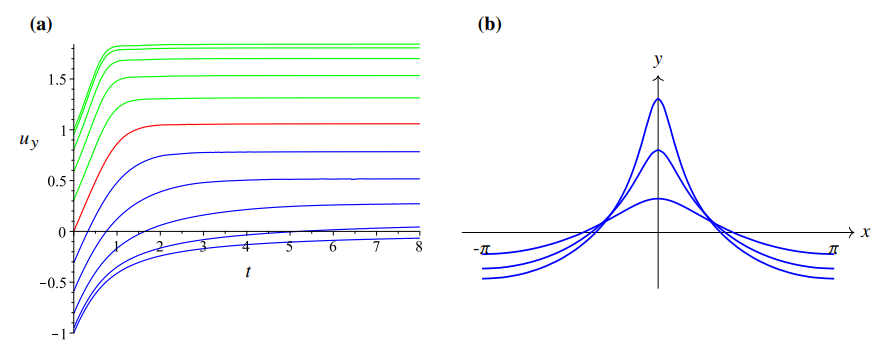}
\caption{({\bf a}): Numerically computed vertical speed of 11 elements initially evenly 
spaced along the surface ($y=0$) from $x=-\pi$ (initial speed $-1$) to $x=0$ 
(initial speed $1$). $\epsilon=0$ throughout.  ({\bf b}): Fluid surface at $t=0.5$, $t=1$ and $t=1.5$.}
\label{fig: 2}
\end{figure}

\begin{figure}
\centering
\includegraphics[width=7cm]{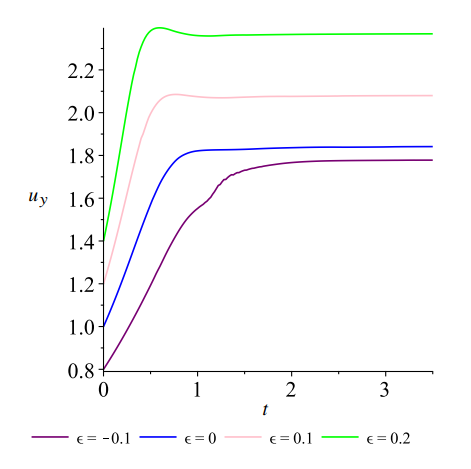}         
\caption{Tip velocity as a function of time for different values of $\epsilon$.}
\label{fig: 3}
\end{figure}

Once the jet starts to extend out of the body of the fluid, all fluid elements 
in the jet end up close to the surface. As a consequence, pressure gradients 
must rapidly become negligible, implying that all fluid elements in the jet move ballistically. 
 This means that, in the long-time limit, an element which reaches a height $y$ above the 
 fluid surface at time $t$ has a vertical speed given approximately by
 \begin{equation}
 \label{eq: 2.13}
 u_y\sim \frac{y-y_0}{t}
 \ .
 \end{equation}
 Figure \ref{fig: 2}({\bf a}) shows the numerically computed vertical speed of different fluid elements 
 on the surface as a function of time, for $\epsilon=0$. The elements close to the tip (top curve)
 rapidly approach a constant vertical speed. Those elements which are close to the trough 
 have a speed which decays slowly, as $1/t$, in accord with (\ref{eq: 1.6}). 
 Figure \ref{fig: 2}({\bf b}) shows the form of the surface at three values of the time 
 $t$, showing that the surface has undergone a significant deformation by the time 
 the tip reaches its maximum speed. This suggests that it would be difficult to make an
 analytical estimate for $v^\ast$ and therefore we use the numerically determined 
 value of $v^\ast=1.84$ for $\epsilon=0$. 
 Figure \ref{fig: 3} shows the time-dependence of the speed of the tip, $x=0$, for different 
 values of $\epsilon$.
 
Section \ref{sec: 3} will describe how equations (\ref{eq: 2.5}) and (\ref{eq: 2.6}) can be 
used, together with (\ref{eq: 2.13}), to deduce the hyperbolic form of the body of the jet,
described by equation (\ref{eq: 1.4}). Here we use the equations of motion for material points 
on the boundary, expressed in the $(u,v)$ coordinate system. In this coordinate system, 
the curve representing the surface extends very rapidly along the positive $u$ 
axis as $t\to \infty$: if (as expected) $y\sim v^\ast t$ then $r\sim \exp(v^\ast t)$. This suggests that 
the closed curve representing the surface has a much greater extension along the $u$-axis
than along the $v$-axis. This would have the benefit of allowing us to 
use a \lq slender-body' approach to determining the potential.

\begin{figure}
\centering
\includegraphics[width=13cm]{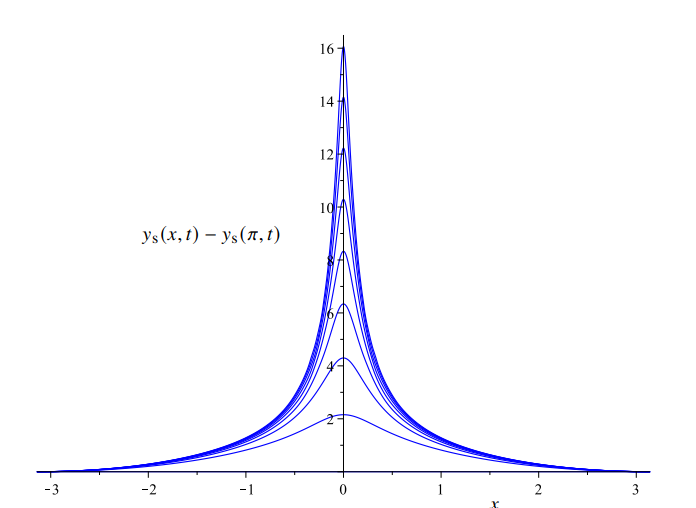}  
\caption{Fluid boundary at $t=0,1,...,8$ from numeric calculations, vertically shifted so that $y=0$ at $x=\pm \pi$ at all times.}
\label{fig: 4}
\end{figure}

Our results indicate that the base of the jet does not get narrower as time increases, rather it approaches a 
time-independent profile, which simply sinks lower as time increases, in accord with 
equation (\ref{eq: 1.6}). Figure (\ref{fig: 4}) shows evidence of this from our numerical model.

\section{Form of body of jet}
\label{sec: 3}

The form of equation (\ref{eq: 1.1}) suggests considering solutions 
of the form (\ref{eq: 2.12}), because the time derivative and the 
gradient-squared term both lead to factors of $t^{-2}$. Furthermore, for reasons 
discussed in sub-section \ref{sec: 2.5} above, we 
assume that the boundary is slender, being extended in the $u$-direction.
A solution of the Laplace equation which is valid close to the $u$-axis is is then well approximated by 
\begin{equation}
\label{eq: 3.1}
\phi(u,v,t)\sim \frac{1}{t}\left[g(u)-\frac{1}{2}g''(u)v^2\right]
\end{equation}
for some function $g(u)$. For the region of the $(u,v)$ plane close to the $u$-axis we have
\begin{equation}
\label{eq: 3.2}
v\sim \theta u
\ .
\end{equation}
Now we introduce the pivotal assumption, namely that \emph{at large time, the 
material points on the boundary move tangentially along this curve}. This curve is defined 
by a function $f$:
\begin{equation}
\label{eq: 3.3}
v=f(u)
\ .
\end{equation}
Now use equations (\ref{eq: 2.5}) to determine a differential equation for $f(u)$:
\begin{equation}
\label{eq: 3.4}
f'=\frac{\frac{{\rm D}v}{{\rm D}t}}{\frac{{\rm D}u}{{\rm D}t}}
=\frac{\frac{\partial \phi}{\partial v}}{\frac{\partial \phi}{\partial u}}=-\frac{g''(u)}{g'(u)}f
\end{equation}
and integrating gives
\begin{equation}
\label{eq: 3.7}
f=\frac{C}{g'}
\ .
\end{equation}
In order to determine the form of the jet, we need to identify the function $g(u)$.
Our \lq slender body' assumption means $v$ is negligible relative to $u$, so we consider $v=0$. 
We can infer that, close to $x=0$ and when $t$ is large, $\phi\sim y^2/2t$, because 
this gives a vertical velocity $u_y=\partial \phi/\partial y=y/t$ in accord with (\ref{eq: 2.13}). 
A more refined estimate of $\phi$ can be obtained from substituting $\phi(u,0,t)=\frac{1}{t}g(u)$ into
(\ref{eq: 2.6}), leading to
\begin{equation}
\label{eq: 3.8}
g(u)=\frac{1}{2}(ug^\prime(u))^2
\ .
\end{equation}
Hence
\begin{equation}
\label{eq: 3.9}
g(u)=\frac{(\ln(\beta u)^2}{2}
\end{equation}
where $\beta$ is a constant of integration.
Hence
\begin{equation}
\label{eq: 3.10}
f(u)=C\frac{u}{\ln(\beta u)}
\ .
\end{equation}
Using the fact that the profile must be symmetric about $x=0$, and mapping back to the $xy$-plane using 
$x = -\theta = -\arctan(f(u)/u)$ and $y = \ln(r) = \ln(\sqrt{u^2 + f(u)^2})$ gives, for large $u$
\begin{equation}
\label{eq: 3.11}
x =\pm \arctan\left(\frac{C}{\ln(\beta u)}\right) \approx \pm \frac{C}{\ln(\beta u)}
\end{equation}
\begin{equation}
\label{eq: 3.12}
y = \ln(u) + \ln \left( \sqrt{1 + \frac{C^2}{\ln(\beta u)^2}} \right) \approx \ln(u)
\ .
\end{equation}
We conclude that, when $u\gg1$ (implying that $y$ is large),
\begin{equation}
\label{eq: 3.13}
x \sim \pm \frac {C}{y+\ln(\beta)}
\end{equation}
a hyperbola, close to the $x$-axis. Note that, because we assumed 
that the trajectories are tangent to the boundary curve at any instant 
in time, $\beta$ may be time-dependent. Equation (\ref{eq: 3.13}) is then 
consistent with (\ref{eq: 1.6}). The fact that we were able to find a 
solution validates our assumption that the velocity is tangent to the surface. Comparison 
with equation (\ref{eq: 1.5}) indicates that we should choose $\beta$ so that
$|x|$ becomes large when $y\to B-D\ln t$, that is as $t\to \infty$ the form of the 
boundary at large $y$ is
\begin{equation}
\label{eq: 3.14}
x \sim \pm \frac {C}{y-B+D\ln t}
\end{equation}
where other constants, $B$, $C$, $D$, are still to be determined.

\begin{figure}
    \centering
    \includegraphics[width=14cm]{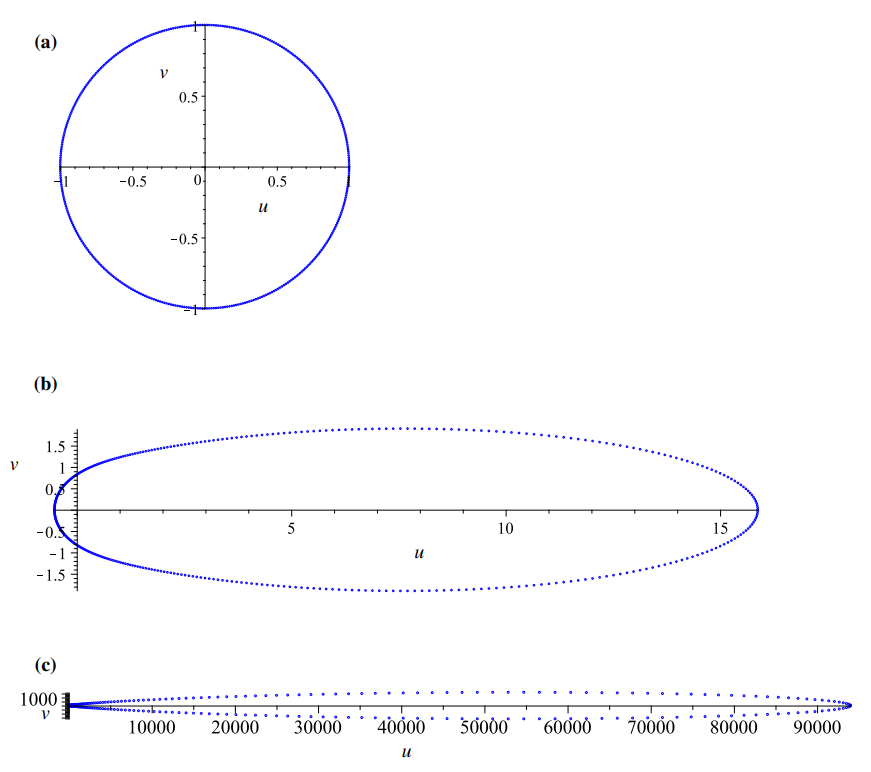}  
\caption{The fluid surface is represented by a closed curve in $(u,v)$ space, 
which starts from a unit circle at $t=0$, and which rapidly extends along the positive
$u$-axis Panel (a) is at $t=0$, panel (b) at $t=1.6$ and panel (c) at $t=6$. 
The large aspect ratio as $t\to \infty$ validates the use of a slender-body 
approximation.}
\label{fig: 5}
\end{figure}

Figure \ref{fig: 5} illustrates the numerically computed surface in the $(u,v)$ plane,
showing evolution from a unit circle at $t=0$ to a curve which is extended along 
the positive $u$-axis. The large aspect ratio of this curve justifies the use of a 
\lq slender-body' approach. This figure was produced by calculating the boundary 
in $(x,y)$ coordinates using the methods described in Appendix A, and using (\ref{eq: 2.2})
to transform to the $(u,v)$ plane. Solving equations (\ref{eq: 2.5}), (\ref{eq: 2.6}) numerically
presents a challenge, because of the explosive growth of the boundary curve when expressed
in these coordinates. We were able to do this to just beyond $t=1$ using a different approach 
to the calculations than our principal method. Up to the point of failure the two methods 
show excellent agreement. Details are provided in Appendix A.

\section{Relation between tip speed and width parameter}
\label{sec: 4}

In section \ref{sec: 3} it was argued that the width of the jet is, to leading order 
as $t\to \infty$, $\Delta x\sim 2C/y$ for some coefficient $C$, which is independent of time.
The tip advances with a speed $v^*$. The values of $C$ and $v^*$ are related by conservation 
of energy. 

The kinetic energy is given by equation (\ref{eq: 2.7}).
Most of the kinetic energy is contained in the region close to the tip,
where the vertical speed is nearly independent of $x$ within the jet.
If the width of the jet is $\Delta x(y)$, then the 
energy is approximated by
\begin{equation}
\label{eq: 4.1}
E=\frac{1}{2}\int^{v^\ast t+y_0}_{y_{\rm b}} {\rm d}y\ \Delta x(y) [u_y(y)]^2
\end{equation}
where $u_y(y)$ is the vertical speed at height $y$, and $y_{\rm b}$ a value close to the base of the jet. 

We know that $u_y(y)=(y-y_0)/t$ (see equation (\ref{eq: 2.13})).  
Close to the tip of the jet, $y_{\rm s}$ may be assumed to be a quadratic 
function of $x$, and as we move away from the tip region, $y_{\rm s}$ may be approximated by 
a hyperbola. We can, therefore, replace the profile described by (\ref{eq: 3.14}))
by a function of the form
\begin{equation}
\label{eq: 4.2}
y(x,t)\sim Y+\frac{C}{\sqrt{x^2+x_0^2}}
\end{equation}
where $Y=B-D\ln t$ and where $x_0(t)$ will be chosen so that $y=y_0+v^\ast t$ when $x=0$. 
To leading order as $t\to\infty$, we have $x_0\sim C/(v^\ast t)$.

\begin{figure}
\centering
\includegraphics[width=14cm]{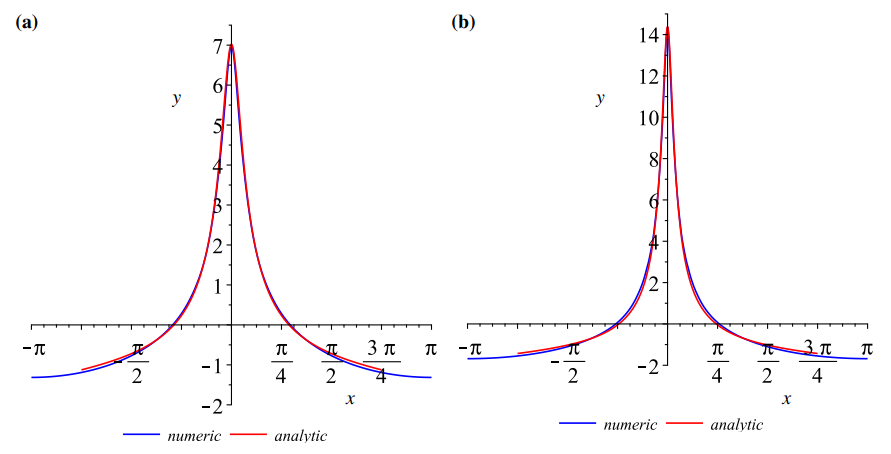}          
\caption{Numerically determined profile of the tip of the 
jet, described by equation (\ref{eq: 4.2a}), using the coefficient $C$ estimated 
using (\ref{eq: 4.4}), (for the case $\epsilon=0$):  ({\bf a}) $t=4$, ({\bf b}) $t=8$.}
\label{fig: 6}
\end{figure}

Note that the energy integral (\ref{eq: 4.1}) is dominated by contributions 
from the tip, rather than the base, because the velocity approaches
zero at the base. The energy is then, to leading order as $t\to \infty$,
\begin{equation}
\label{eq: 4.3}
E=C\int_{y_{\rm b}-y_0}^{v^\ast t}{\rm d}y'\ 
\frac{y'}{v^\ast t^3}\sqrt{(v^{\ast}t)^2-y'^2}\sim \frac{C v^{\ast 2}}{3}
\ .
\end{equation}
Knowing $E$, we can determine $C$ in terms of $v^\ast$:
\begin{equation}
\label{eq: 4.4}
C=\frac{3\pi(1+2\epsilon^2)}{2v^{\ast 2}}
\ .
\end{equation}
Using a precise expression for $x_0$, for which (\ref{eq: 4.2}) gives $y=y_0+v^\ast t$
when $x=0$, we obtain our best approximation for the form of the jet:
\begin{equation}
\label{eq: 4.2a}
y(x,t) \sim B-D\ln t+\frac{C}{\sqrt{x^2+\frac{C^2}{(y_0+v^\ast t-B+D\ln t)^2}}}
\ .
\end{equation}
Figure \ref{fig: 6} compares the numerical computed tip profile with 
the prediction of equations (\ref{eq: 4.2a}) and (\ref{eq: 4.4}), for the case 
$\epsilon=0$. The fitted parameters were $v^\ast=1.84$, $y_0=-0.34$, $B=-0.80$.

\section{Modelling the base of the jet}
\label{sec: 5}

\subsection{Lowest level of trough}
\label{sec: 5.1}

First let us consider the height of the lowest point, $x=\pm \pi$, 
at time $t$. This will be estimated by assuming the at base of the jet 
has a fixed shape, and simply sinks in the limit as $t\to \infty$.
The sinking of the lowest point is then determined using conservation 
of volume.

\begin{figure}
\centering
\includegraphics[width=14cm]{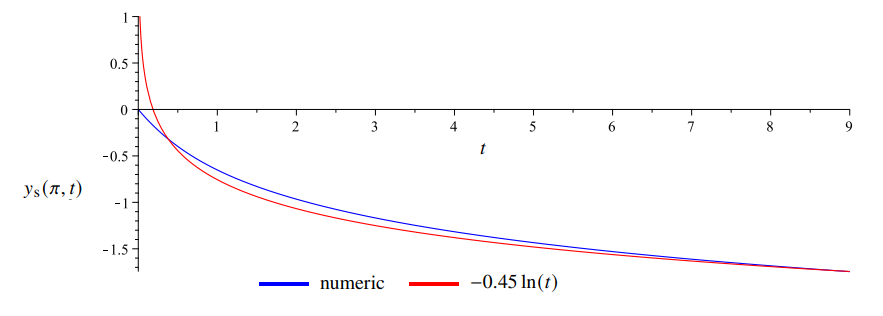}
\caption{
Numerically determined level of the base of the trough as a function of time, compared 
with theory, equations (\ref{eq: 1.5}) and (\ref{eq: 5.2}), which predicts 
$D=0.443\ldots $ (for the case $\epsilon=0$).
}
\label{fig: 7}
\end{figure}

Using equation (\ref{eq: 4.2}) to describe the shape of the tip of the jet,
the volume of the tip above $y=y_{\rm b}$ (with $y_{\rm b}\ll v^\ast t$), is
\begin{equation}
\label{eq: 5.1}
{\cal V}_+ = \int_{y_{\rm b}}^{v^\ast t+y_0}{\rm d}y\ \Delta x(y) \sim 2C
\int_{y_{\rm b}-y_0}^{v^\ast t}{\rm d}y'\ \frac{\sqrt{v^{\ast 2}t^2-y'^2}}{y'v^\ast t}
\sim C \left[\frac{(y_{\rm b}-y_0)^2}{(v^{\ast}t)^2}-2-2\ln\left(\frac{y_{\rm b}-y_0}{2v^\ast t}\right)\right]
\ .
\end{equation}
Note that, in the limit as $t\to \infty$, the term proportional to $t^{-2}$ becomes 
negligible, so that the dependence upon $t$ is logarithmic, with coefficient $2C$.
The increase in ${\cal V}_+$ is compensated by lowering the base region, for which the 
profile is described by equation (\ref{eq: 1.6}), except in a narrowing interval in 
the vicinity of the tip, at $x=0$. Accounting for the fact that the width of the region 
is $2\pi$, the contribution to the area change from the base has a logarithmic term,
${\cal V}_-=-2\pi D\ln t$, where $D$ is the coefficient in (\ref{eq: 1.6}). Hence conservation 
of volume indicates that 
\begin{equation}
\label{eq: 5.2}
D=\frac{C}{\pi}=\frac{3(1+2\epsilon^2)}{2v^{*2}}
\ .
\end{equation}
Figure \ref{fig: 7} shows the numerically determined level of the base of the trough, compared 
with a fit using the theoretically estimated value of the coefficient $D$ in (\ref{eq: 1.6}).

\subsection{Shape of base of trough}
\label{sec: 5.2}

Consider the form of the jet when expressed in terms of the 
$(u,v)$ variables (defined by (\ref{eq: 2.2})). The bottom of the trough 
at $x=\pm \pi$ corresponds to a point in the $(u,v)$ plane, with coordinate 
which approaches the origin as $t\to \infty$: taking the exponential of (\ref{eq: 1.6})
suggests that the position $(u^\ast(t),0)$ corresponding to the bottom of the trough
has
\begin{equation}
\label{eq: 5.3}
u^\ast(t)=-b t^{-D}
\end{equation}
(with $B=\ln b$). In our numerical work we set the value of $b$ to 0.45, using the largest value of $t$ 
(approximately $t=8$) for which our numerical solution was stable. 

Let us consider the consequences of the surface corresponding to a curve in 
the $(u,v)$ plane scaling in the same manner as (\ref{eq: 5.3}). Define scaled
coordinates $(U,V)$, a transformed time variable $\tau$, and a transformed potential $\Psi(U,V)$ 
by writing 
\begin{equation}
\label{eq: 5.4}
(U,V)=t^{D}(u,v)
\ ,\ \ \ 
\tau=\ln(t)
\ ,\ \ \ 
\phi(u,v,t)=\frac{1}{t}\Psi(U,V,\tau)
\ .
\end{equation}
In these variables, the Longuet-Higgins and Cokelet equations of motion, 
(\ref{eq: 2.5}), (\ref{eq: 2.6}) become
\begin{eqnarray}
\label{eq: 5.5}
\frac{{\rm D}U}{{\rm D}\tau}&=&(U^2+V^2)\frac{\partial \Psi}{\partial U}+DU
\nonumber \\
\frac{{\rm D}V}{{\rm D}\tau}&=&(U^2+V^2)\frac{\partial \Psi}{\partial V}+DV
\end{eqnarray}
and
\begin{equation}
\label{eq: 5.6}
\frac{{\rm D}\Psi}{{\rm D}\tau}=\frac{(U^2+V^2)}{2}
\left[\left(\frac{\partial \Psi}{\partial U}\right)^2+\left(\frac{\partial \Psi}{\partial V}\right)^2\right] +\Psi
\ .
\end{equation}
We hypothesise that the boundary approaches a fixed curve when expressed 
in terms of the $(U,V)$ variables. Note that, if  $\Psi(U,V,\tau)$ is independent of $\tau$, then 
equation (\ref{eq: 5.5}) describes motion in a fixed velocity field. The boundary curve 
might be represented by a curve representing the unstable manifold attached to an 
unstable fixed point at $U^\ast=-b$.

The curve in $(U,V)$ space represents a family of curves in $(x,y)$ space, corresponding 
to different values of $t$. Consider how the value of $y$ changes for fixed $x$, as $t$ 
is varied. According to (\ref{eq: 2.2}), fixing $x$ corresponds to fixing the polar angle $\theta $
in the $(u,v)$ plane, and hence also in the $(U,V)$ plane. The ray at polar angle intersects the curve 
at a point $(U(\theta),V(\theta))$, with radius $R(\theta)=\sqrt{U^2+V^2}$. The corresponding 
value of $y$ is $y=\ln(t^{-D}R(\theta))$. It follows that the values of $y$ are shifted by 
the same amount, independent of $x$, as $t$ increases.

The unstable fixed point at $U^\ast$ corresponds to the unstable fixed point 
of the surface flow at $x=\pm \pi$. We can expand $\Psi(U,V)$ about this fixed 
point, and determine coefficients by comparing with the unstable fixed point of the 
surface flow. Noting that $\Psi$ should be a harmonic function, to leading order
\begin{equation}
\label{eq: 5.7}
\Psi(U,V)=\frac{\mu}{2}(U^2-V^2)+\nu U
\end{equation}
where $\mu$. $\nu$ are constants to be determined. 
To determine the fixed point $U^\ast$=-$b$, set ${\rm D}U/{\rm D}\tau=0$ to obtain
\begin{equation}
\label{eq: 5.8}
\mu b^2-\nu b+D=0
\ .
\end{equation}
The equation of motion for $V$ in the vicinity of the fixed point satisfies
\begin{equation}
\label{eq: 5.9}
\frac{{\rm D}V}{{\rm D}\tau}\sim \left[D-\mu b^2\right]V
\end{equation}
so that 
\begin{equation}
\label{eq: 5.10}
V(\tau)\sim V_0 \exp[\lambda \tau]=V_0 t^\lambda
\ ,\ \ \ 
\lambda=D-\mu b^2
\ .
\end{equation}
The principle discussed in subsection \ref{sec: 2.3} above implies that 
material points on the surface move with an approximately fixed velocity
while they are still in the vicinity of the unstable fixed point of the surface flow.
If their initial displacement from this fixed point is $\delta x_0$, the displacement
after time $t$ is
\begin{equation}
\label{eq: 5.11}
\delta x(t)=\delta x_0+ (1-4\epsilon)\delta x_0 t
\ .
\end{equation}
Because $\delta x=\pi-\theta$, and $\pi-\theta\sim -V/U$ close to the negative 
$U$ axis, equations (\ref{eq: 5.10}) and (\ref{eq: 5.11}) are only consistent
as $t\to \infty$ if 
\begin{equation}
\label{eq: 5.12}
\lambda=D-\mu b^2=1
\ .
\end{equation}
We have assumed that $b$ and $D$ are known. Equations (\ref{eq: 5.8}) 
and (\ref{eq: 5.12}) then enable us to determine $\mu$ and $\nu$. Close to
the fixed point, $U^\ast=-b$, the equation of motion for $U$ is approximated by 
\begin{equation}
\label{eq: 5.13}
\frac{{\rm D}U}{{\rm D}\tau}\sim \frac{D}{b}V^2
\end{equation}
so, using equation (\ref{eq: 5.10}) with $\lambda=1$ and as $V=0$ 
at $U=-b$, then the curve in the $(U,V)$ plane is approximated by
\begin{equation}
\label{eq: 5.14}
V_{\rm b}(U)\sim \pm \sqrt{\frac{2b}{D}(U+b)}
\ ,\ \ \ 
U+b>0
\end{equation}

\begin{figure}
\centering
\includegraphics[width=9cm]{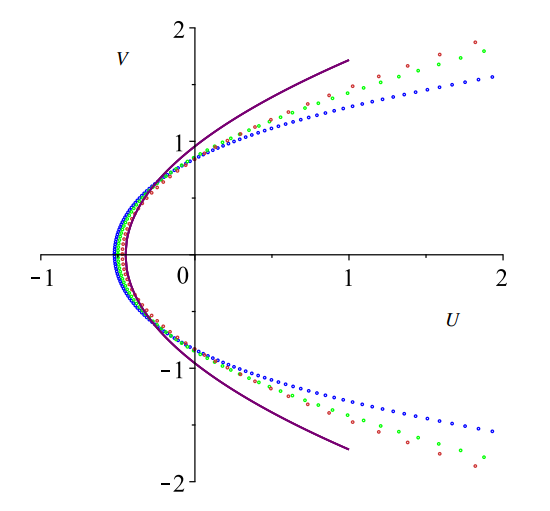}
\caption{Numerically determined surface contour in $(U,V)$ 
space at $t=2$ (blue), $t=4$ (green) and $t=8$ (orange), using the value of $D$ predicted 
by (\ref{eq: 5.2}), and $b=0.45$. The figure also shows the analytic 
approximation (purple) to this limiting curve, equation (\ref{eq: 5.14}). 
These data are for $\epsilon=0$.}
\label{fig: 8}
\end{figure}

Figure \ref{fig: 8} shows the numerically determined surface contour in $(U,V)$ 
space, at $t=2,4,8$, using the value of $D$ predicted by (\ref{eq: 5.2}). 
The curves are nearly coincident, consistent with the expectation that they should 
approach a limiting curve as $t\to \infty$. The figure also shows the analytic 
approximation to this limiting curve, equation (\ref{eq: 5.14}). Close to the critical point, 
$(U,V)=(U^\ast,0)$, the numerical curves approach this parabolic approximation 
as $t$ increases. Away from the critical point, there is a an increasing discrepancy, 
which is already significant when $U=0$.
It would be possible in principle to continue the expansion about the 
fixed point to higher orders in $V$, but we believe that this leading order 
solution is sufficient to understand the physical principles.

\begin{figure}
\centering
\includegraphics[width=9cm]{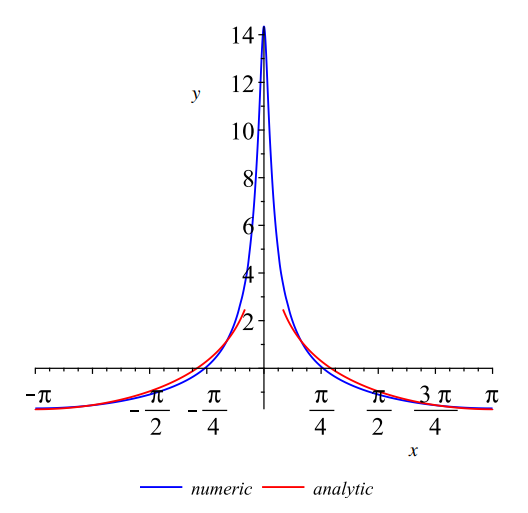}
\caption{Numerically determined level of the base of the trough at $t=8$, compared 
with theory, equation (\ref{eq: 5.2}), which predicts $D=0.443\ldots $ (for the case $\epsilon=0$)
We set $b=0.45$ by fitting (\ref{eq: 5.3}) to the numerical solution at $t=8$.}
\label{fig: 8.5}
\end{figure}

Equation (\ref{eq: 5.14}) expresses the form of the boundary close to the 
trough in the $(U,V)$ plane. We can use (\ref{eq: 2.2}) and (\ref{eq: 5.4}) to 
transform this to a curve in the original Cartesian coordinates. The result, illustrated 
in figure \ref{fig: 8.5}, is in good agreement with numerical simulations of the boundary.

\section{Discussion}
\label{sec: 6}

\subsection{Synthesis}
\label{sec: 6.1}

\begin{figure}
\centering 
\includegraphics[width=14cm]{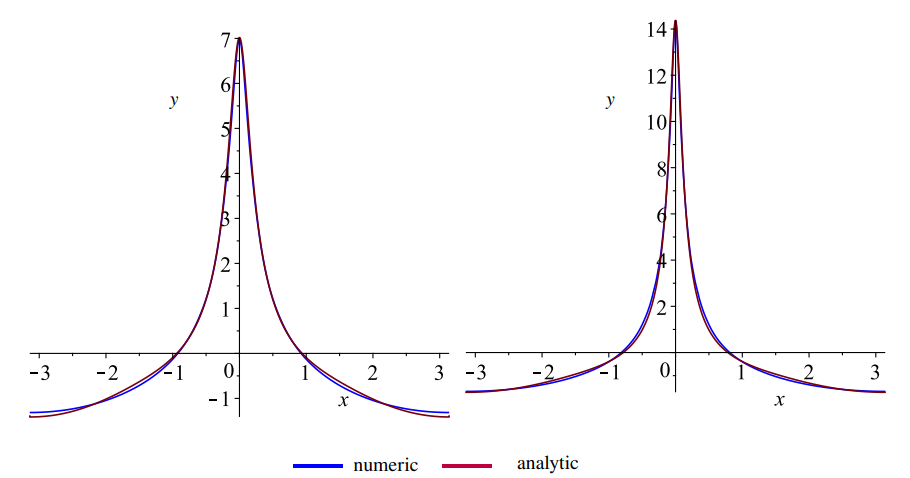}
\caption{Frames (a) and (b) compare the numerically determined surface contour in $(x,y)$ 
space, at $t=4,8$ respectively, with theory (equations (\ref{eq: 4.2a}), (\ref{eq: 4.4}) describing the jet, 
and (\ref{eq: 5.14}), describing the base, interpolated according to equations 
(\ref{eq: 6.1}) and (\ref{eq: 6.2})). These data are for $\epsilon=0$. }
\label{fig: 9}
\end{figure}

We have obtained two approximations for the form of the surface, 
one valid in the jet (equations (\ref{eq: 4.2a}), (\ref{eq: 4.4})), and the other
describing the base of the jet (equation (\ref{eq: 5.14})). Figures \ref{fig: 6}
and \ref{fig: 8.5} show that these approximations work well, close to $x=0$ 
and $x=\pm \pi$ respectively. Figure \ref{fig: 9} compares the numerically 
computed profile with an interpolation between these two expressions:
\begin{equation}
\label{eq: 6.1}
y(x,t)=w(x)\,y_{\rm j}(x,t)+[1-w(x)]\,y_{\rm b}(x,t)
\end{equation}
where $y_{\rm j}(x,t)$ and $y_{\rm b}(x,t)$ are the profiles for the jet and base 
regions, equations (\ref{eq: 4.2a}) and (\ref{eq: 5.14}) respectively, and where the
interpolating weight function was 
\begin{equation}
\label{eq: 6.2}
w(x)=\left\{
\begin{array}{cc}
-2x^2/\pi^2+1 & 0 \le |x| < \pi/2  \cr
2x^2/\pi^2-4|x|/\pi+2  &  \pi/2 \le |x| \le \pi \cr
\end{array}
\right.
\ .
\end{equation}
There is excellent agreement between the interpolated theoretical function 
and the numerically determined surface contour, for $t=4$ and $t=8$. 

\subsection{Conclusions}
\label{sec: 6.2}

Understanding the ejection of jets from the surface of a simple fluid is a fundamental
and surprisingly challenging problem. An ellipsoidal body subjected to a simple 
extensional flow was solved by \cite{Dir60}, see also \cite{Lon72}. 
In this paper we considered what is 
arguably the simplest extension of that model: an initially flat surface of a deep
body of liquid in two dimensions is subject to a spatially periodic, impulsive disturbance. 
Even for this case, it is difficult to produce numerical results which are reliable
at large time, and it appears to be impossible to find a precise asymptotic solution 
which describes all regions of the motion. However, we were able to 
show, in section \ref{sec: 3}, that away from the base and tip, jet profile is hyperbolic. 
This is at variance with an earlier suggestion \cite{Lon01a} that the profile 
is a cusp.

We also found that the base region moves downwards without changing its form, and that the 
depth is logarithmic in time. There is no separatrix, dividing material elements which will 
eventually enter the jet from those which remain in the bulk. Eventually, every 
fluid element, at any initial depth, becomes part of the jet.

The speed $v^\ast$ of the tip of the jet must be determined numerically, 
but once this has been determined, we can predict the evolution of the jet at large time. 
In particular, the parameters $C$ determining the width of the jet, and $D$ 
describing the depth of the trough, are given by explicit formulae, (\ref{eq: 4.4}) and 
(\ref{eq: 5.2}). The parameter $b$ introduced in (\ref{eq: 5.3}) is not determined by 
an explicit formula, but it can be determined by the requirement that volume is conserved.

The argument in \cite{Lon01a} uses a somewhat different approach: seeking a coincidence
of two contours (the surface is a zero contour of both the pressure $p$ and its material 
derivative, ${\rm D}p/{\rm D}t$). This is implemented by developing the velocity potential 
as a Taylor series. His approach suffers from the difficulty that the method develops a 
velocity potential which has unbounded growth at infinity, and which cannot be matched 
to any physical boundary condition.

There remain many open questions about the ejection of jets from bodies of inviscid 
fluid. In our case the point of origin of the jet and its direction ($x=0$, upwards) were 
obvious, but in cases with a more general initial condition, the number, origin, direction, 
speed and relative mass of the jets are hard to determine with certainty. Extensions 
to three dimensions are also non-trivial.

\backsection[Supplementary data]{The data that support the findings of this study are 
available upon reasonable request from the authors.}

\backsection[Funding]{This research received no specific grant from any funding agency, 
commercial or not-for-profit sectors. }

\backsection[Declaration of interests]{The authors report no conflict of interest.}

\backsection[Author ORCIDs]{

Andrew Wilkinson https://orcid.org/0009-0002-3159-1037; 

Michael Wilkinson https://orcid.org/0000-0002-5131-9295.}

\appendix

\section{Numerical technique}
\label{sec: A}

\subsection*{Basic method}
\label{sec: A.1}

The velocity potential at time $t$ will be expressed in terms of Fourier coefficients, 
$\alpha_n(t)$:
\begin{equation}
\label{eq: A.1}
\phi(x,y,t)=\sum_{n=0}^{N_c} \alpha_n(t)\cos(nx) \exp(ny)
\end{equation}
so the velocity components are
\begin{eqnarray}
\label{eq: A.2}
u_x(x,y,t)=\frac{\partial \phi}{\partial x}&=&-\sum_n n\,\alpha_n(t)\sin(nx)\exp(ny)
\nonumber \\
u_y(x,y,t)=\frac{\partial \phi}{\partial y}&=&\sum_n n\,\alpha_n(t)\cos(nx)\exp(ny)
\ .
\end{eqnarray}

There are $M$ material points on the boundary, with coordinates and 
potentials $X_i(t)$, $Y_i(t)$, $\Phi_i(t)$. Initially these points are taken to 
be uniformly spaced on the $x$-axis between $-\pi$ and $\pi$. The initial impulse given to
these material points is per equation (\ref{eq: 1.2}), so that when $\epsilon=0$,  
$\alpha_n(0)=\delta_{n,1}$ in (\ref{eq: A.1}). Values used for results presented were 
$N_c=18$ (the number of cosine Fourier components)
and $M=401$ (the number of material points).

Because (\ref{eq: A.1}) is automatically a solution of the Laplace 
equation, there is no need to solve the Laplace equation.
The numerical method used to evolve the motion of the boundary (both the positions 
and potentials of the material points on the boundary) is based on the approach of 
Longuet-Higgins and Cokelet, in which the boundary evolves according to equations 
(\ref{eq: 2.1}). Given that the coefficients $\alpha_n$ are known at some initial time $t$, 
the new state of the boundary (coordinates and potentials) is given by
\begin{eqnarray}
\label{eq: A.3}
X_i^\prime&=&X_i+\left(\frac{\partial\phi}{\partial x}\right)_i\delta t
\ , \ \ \
\
Y_i^\prime=Y_i+\left(\frac{\partial\phi}{\partial y}\right)_i\delta t
\nonumber \\
\Phi_i^\prime&=&\Phi_i+\frac{1}{2}\left(\left(\frac{\partial\phi}{\partial x}\right)_i^2+\left(\frac{\partial\phi}{\partial y}\right)_i^2\right)\delta t
\end{eqnarray}
for $i=1,...,M$ where equations (\ref{eq: A.2}) have been used to evaluate the velocity 
components on the boundary points $X_i,Y_i$. The evolved data (\ref{eq: A.3}) 
are then used to determine the new Fourier representation at time $t+\delta t$:
\begin{equation}
\varphi_i(\{\alpha_n^\prime\})=\sum_{n=0}^{N_c} \alpha_n^\prime\cos(nX_i^\prime) \exp(nY_i^\prime)
\nonumber
\end{equation}
where the new Fourier coefficients $\alpha_n^\prime$ are determined by a least-squares fit of 
$\varphi_i$ to the evolved data $\Phi_i^\prime$, that is by numerical minimisation of the functional
\begin{equation}
\label{eq: A.4}
f(\{\alpha_n^\prime\})\equiv\sum_{i=1}^M(\varphi_i(\{\alpha_n^\prime\})-\Phi_i^\prime)^2
\ ,\ \
\end{equation}
where the earlier values $\alpha_n$ were used by the minimisation routine as the initial guess. 
Once the new coefficients $\alpha_n^\prime$ are determined, the substitutions $\alpha_n^\prime\to\alpha_n$ and 
$(X_i^\prime,Y_i^\prime)\to (X_i,Y_i)$ in equations (\ref{eq: A.1}) and (\ref{eq: A.2})
give the new boundary point potential and velocities. Iteration of this process gives a 
numerical approximation for the boundary height 
function $y_s(x,t)$, although this method was found to become unstable at 
around $t=2$ when the jet has extended significantly beyond the body of fluid, 
and an extension to our method described in the following sub-section, was employed to circumvent this problem.

\subsection*{Two-region parametrisation of boundary}
\label{sec: A.2}

The method for propagating the boundary requires information about 
the solution of the Laplace equation in the interior region.
After the jet extends a significant distance above the surface, 
it is easier to give an accurate representation of $\phi(x,y,t)$ 
if the boundary and the interior are divided into two regions,
which will be referred to as the \emph{base} and the \emph{jet}.  One benefit of this 
scheme is that it extends the time of numerical stability, considerably beyond $t=2$.

The interior potential is determined by a different method for the base
($M_1$ points) and the jet ($M_2$ points). The two sets are defined by 
overlapping threshold values for $Y$, namely $Y_1$, $Y_2$, with $Y_2>Y_1$.
The jet region is defined by $Y>Y_1$ and the base by $Y<Y_2$. 
We found the optimal values of $Y_1$ and $Y_2$ to be $0.01$ and $3.5$ respectively, and 
these are used in the results presented.
The predicted increments of $X_i$, $Y_i$ and $\Phi_i$ are
different for the two regions, and the actual increment that is
applied is an interpolation
\begin{equation}
\label{eq: A.8}
\delta \Phi_i=G\left(\frac{Y_i-Y_1}{Y_2-Y_1}\right)\delta \Phi^{(2)}_i
+\left[
1-G\left(\frac{Y_i-Y_1}{Y_2-Y_1}\right)
\right]\delta \Phi^{(1)}_i
\end{equation}
where $\delta \Phi^{(1)}_i$ and $\delta \Phi^{(2)}_i$ are the increments 
predicted using the base and jet potentials, respectively. For the interpolation function $G(x)$, we used
\begin{equation}
\label{eq: A.9}
G(x)=\left\{
\begin{array}{cc}
0 & x<0 \cr
2x^2  & 0<x<1/2 \cr 
4x-2x^2-1 & 1/2<x<1\cr
1 & x>1
\end{array}
\right.
\ .
\end{equation}
In the initial stages of the motion, the maximum height of the boundary, $Y_{\rm max}$, 
is less than the first threshold, $Y_1$, so all material points on the boundary move according to the base 
potential expressed in the Fourier series, as described in the previous section. When $Y_{\rm max}$ starts to 
exceed $Y_1$ those points with $Y_i>Y_1$ contribute to the jet potential (described below), 
and are mixed in according to (\ref{eq: A.8}).

In the jet region, at any given time, we assume that the field on the central axis, $x=0$, 
is given by a function $F_0(y)$, which can be expressed as 
\begin{equation}
\label{eq: A.12}
F_0(y)=\sum_{k=0}^{N} \phi_k \left(\frac{y-Y_1}{Y_{\rm max}-Y_1}\right)^k
\ .
\end{equation}
By symmetry, away from the axis $\phi(x,y)$ is an even function, 
which we expand as a series
\begin{equation}
\label{eq: A.10}
\phi(x,y)=\sum_{n=0}^{(N-1)/2}F_n(y)x^{2n}
\ .
\end{equation}
Imposing that $\nabla^2\phi=0$, we find $F_1(y)=-F_0''(y)/2$, $F_2(y)=-F_1''(y)/12=F_0''''(y)/24$, etc.,
so that
\begin{equation}
\label{eq: A.11}
\phi(x,y)\approx F_0(y)-\frac{1}{2}F_0''(y)x^2+\frac{1}{24}F_0''''(y)x^4\;+\cdots
\ .
\end{equation}
we do a least-squares fit of (\ref{eq: A.11}) to the given values of $X_i$, $Y_i$ and $\Phi_i$ to determine the \lq jet' coefficients $\{\phi_k\}$. This fit includes only those boundary points with $Y_i>Y_1$. 

Here the coefficients $\phi_k$ may become independent of time in the long-time limit. 
Because the velocity profile becomes linear and we expect $F_0(y)$ is well-approximated 
by a quadratic, so that the $\phi_k$ become small for $k\ge 2$. In fact we did not find 
this to be the case, apparently due to the least-squares fit being 
ill-conditioned. In practice, we obtained the most consistent numerical results 
when we included terms up to $\phi_{9}$ i.e. $N=9$. After approximately $t=9$ the method became 
unstable. The numerical work was coded with \emph{Maple}, and the high-precision 
arithmetic used by that package may have been beneficial.
 
Finally, when $Y_{\rm max}$ exceeds the second threshold, $Y_2$, the least-squares fit 
for the base (\ref{eq: A.4}) includes only those points with $Y_i<Y_2$.

\subsection*{Alternative Delauney/functional method to determine the boundary in the $(u,v)$ plane}
\label{sec: A.3}

In the $(u,v)$ plane our fluid is initially bounded by a unit circle. Here, we also used a completely 
different approach to solving Laplace's equation at each time-step by finding a field $\phi(x,y)$ which 
minimises the functional 
\begin{equation}
\label{eq: A.13}
S=\frac{1}{2}\int {\rm d}x \int {\rm d}y\ (\mbox{\boldmath$\nabla$}\phi)^2
\ .
\end{equation}
To find a tractable form of $S$ we use a Delaunay triangulation of the region containing the 
fluid (initially a unit circle), defining $\phi(x,y)$ by its value on the vertices of each triangle, 
$\phi_i$ (internal vertex, free to vary) and $\psi_i$ (boundary vertex, fixed). The functional is 
then approximated by the quadratic form 
\begin{equation}
\label{eq: A.14}
S=\frac{1}{2}\sum_{ij, {\rm int}} A_{ij} \phi_i \phi_j +\sum_{i, {\rm int}, j, {\rm bdy}} A_{ij}\phi_i\psi_j
+\frac{1}{2}\sum_{ij,{\rm bdy}}A_{ij}\psi_i\psi_j
\end{equation}
where the coefficients $A_{ij}$ are sums of contributions from
each triangle. Minimising $S$ is equivalent to solving a system of linear equations, 
$\sum_{j,{\rm int}} A_{ij} \phi_j+\sum_{j,{\rm bdy}}A_{ij}\psi_j=0$, to determine the internal
potentials $\phi_j$. These then 
allow us to calculate gradients for the boundary points, which can then be advanced by one time-step. 
The process is then repeated, beginning with a new Delaunay triangulation of the new fluid-filled region, for each time-step. 

\begin{figure}
\centering
\includegraphics[width=14cm]{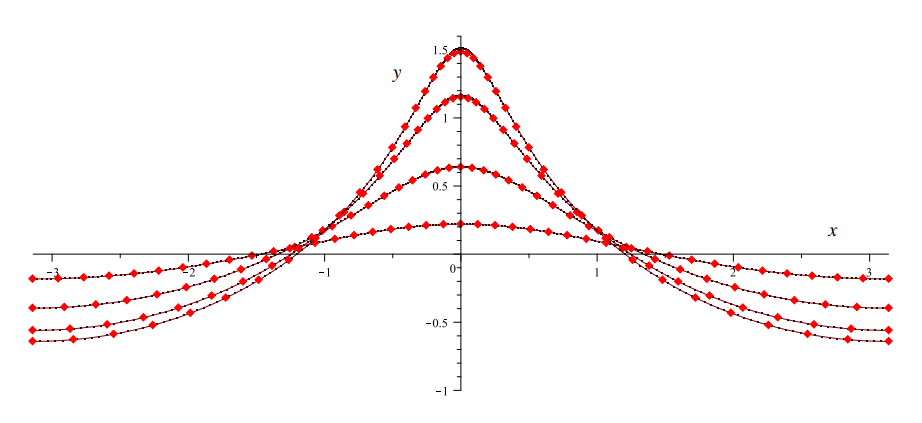}
\caption{
Boundary at intervals up to $t=1$ as determined numerically by the \lq Fourier series' method (black line) and the \lq Delaunay / Functional'  method (red diamonds).
}
\label{fig: 11}
\end{figure}

As mentioned earlier, the $u$ values in the jet tip region grow extremely rapidly in this plane, and the 
calculation exceeds the capabilities of a standard computer after a relatively short time. We managed to 
project the boundary to just over $t=1$, much less than the time managed by our  principal method 
described earlier in this appendix. Nevertheless, the boundary predicted by these two very different 
methods agrees, up to the point where the Delauney/functional method fails. 
Figure (\ref{fig: 11}) shows the boundaries up to $t=1$ as predicted by both numeric 
methods with the Delaunay / functional method results transformed back to the $(x,y)$ plane.

\section{Initial pressure pulse}
\label{sec: B}

Here we determine the initial pressure $p(x,y,0)$.
Throughout the fluid the pressure satisfies
\begin{equation}
\label{eq: 4.5}
\frac{\partial \phi}{\partial t}+\frac{1}{2}(\nabla \phi)^2+p(x,y,t)=\Phi(t)
\end{equation}
and the potential is 
\begin{equation}
\label{eq: 4.6}
\phi(x,y,t)=\sum_{n=0}^\infty \alpha_n(t)\cos(nx)\exp(ny)
\end{equation}
with the only non-zero coefficients at $t=0$ being $\alpha_1(0)=1$, $\alpha_2(0)=\epsilon$, 
and possibly $\alpha_0$. At $t=0$:
\begin{equation}
\label{eq: 4.7}
(\nabla \phi)^2=\exp(2y)+4\epsilon\cos(x)\exp(3y)+4\epsilon^2\exp(4y)
\end{equation}
Because $p=0$ at the surface (initially $y=0$), equations (\ref{eq: 4.6}) and 
(\ref{eq: 4.7})  can only be consistent with (\ref{eq: 4.5}) if $\dot \alpha_1(0)$ is non-zero 
(while $\dot \alpha_n(0)=0$ for $n\ge 2$). Hence we find 
\begin{equation}
\label{eq: 4.8}
p(x,y,0)=\frac{1}{2}\left[1+4\epsilon^2-\exp(2y)-4\epsilon^2\exp(4y)\right]
+2\epsilon\exp(y)\cos(x)\left[1-\exp(2y)\right]
\ .
\end{equation}

After the initial impulse which sets the fluid in motion, 
the pulse of pressure initially accelerates all of the surface 
elements upwards, as is evident from figures \ref{fig: 2} 
and \ref{fig: 3}. However, we were not able to find any relation between 
the tip velocities in figure \ref{fig: 3} and gradients of the initial pressure field.

\end{document}